\documentstyle[11pt,twoside,newpasp]{article}

\input{tcilatex}
\begin{document}

\title{Binary Pulsar Tests of General Relativity in the Presence of Low-Frequency
Noise}
\author{Sergei M. Kopeikin$^{1,2}$ and Vladimir A. Potapov$^{3}$}

\begin{abstract}
The influence of the low-frequency timing noise on the precision of
measurements of the Keplerian and post-Keplerian orbital parameters in
binary pulsars is studied. Fundamental limits on the accuracy of tests of
alternative theories of gravity in the strong-field regime are established.
The gravitational low-frequency timing noise formed by an ensemble of binary
stars is briefly discussed.
\end{abstract}

\affil{$^{1}$ Dept. of Physics \& Astronomy, \newline
University of Missouri-Columbia,\newline
Columbia, MO65211, USA\newline
$^{2}$ On leave: ASC FIAN, Leninskii Prospect 53, Moscow, Russia \newline
$^{3}$ PRAO, ASC FIAN, Leninskii Prospect 53, Moscow, Russia}

\section{Background}

The discovery of the first binary pulsar in 1974 by R. Hulse and J.H. Taylor
had opened fascinating new opportunities for testing alternative theories of
gravitation in the radiative and in the strong-gravitational-field regimes
outside the boundaries of the Solar system. Presently, two distinguished
binary pulsars - PSR B1913+16 and PSR B1534+12 - serve as primary
extra-solar astrophysical laboratories for gravitational physisists. The
radiative $\dot{\omega}-\gamma -\dot{P}_{b}$ test of General Relativity (GR)
in the case of PSR B1913+16 has been performed with a precision about 0.4\%
(Taylor \& Weisberg 1989, Damour \& Taylor 1991). In addition, the binary
pulsars PSR B1534+12 strong-field $\dot{\omega}-\gamma -s$ test of GR has
been done with the precision 1\% (Stairs et al. 1998). What is worth noting
is that the binary pulsar tests of GR have been carried out with an accuracy
being comparable to, or in some aspects, higher than the GR tests in the
Solar system which have been always performed in the near, weak-field zone
of the solar system gravitating bodies.

Reflecting about the future of gravitational experiments in binary pulsars
one can ask about whether the accuracy of GR tests can be further improved
and, if not, what physical influences can limit this accuracy. It is well
known there are two main obstacles for improving precision of any experiment
- systematic and random errors\footnote{%
There exist a third reason which restricts the precision of GR tests in
binary pulsars. It relates to the fact that some additional (''hidden'')
parameters appear in the form of additive linear corrections to measured
values of the main relativistic effects. These parameters can not be
effectively separated from the parameteres making the biggest contribution
to the effect so that one can not improve its numerical value (for more
details see, for example, Damour \& Taylor 1991, Kopeikin 1994, 1996, Wex \&
Kopeikin 1999).}. Systematic errors cause the bias in the mean values of
estimated parameters and can lead to inconsistent conclusions on the
validity of the theory of gravitation under investigation. Fortunately,
continuous improvements being made in observational techniques and growing
knowledge on dynamics of our galaxy can siginificantly reduce or even
eliminate the influence of systematic errors (Damour \& Taylor 1991). On the
other hand, random errors fully determine the magnitude of the numerical
value of the dispersion of measured parameters which itself crucially
depends on the nature of noise in the timing measurements. Usually, on
rather short time spans the white noise of measurement errors dominates.
However, to improve the accuracy of the GR tests one has to observe binary
pulsars over time intervals of about 10 years and longer when the white
noise is strongly supressed and low-frequency (or ''red'') noise dominates
in observed times-of-arrivals (TOA) of pulsar' pulses. The red noise worsens
estimates of measured parameters and increases the errors in their mean
values. What kind of problems will one meet in processing such timing
observations contaminated with the presence of the red noise ? Before trying
to answer this question let us briefly descibe the timing model of binary
pulsars and the properties of red noise.

\section{Timing Model}

The conventional timing model is based on the Damour-Deruelle analytic
parameterization of the two-body problem (Damour \& Deruelle 1985, see also
Klioner \& Kopeikin 1994). Schematically it reads as follows 
\begin{equation}
N(T)=N_{0}+\nu T+\frac{1}{2}\dot{\nu}T^{2}+\epsilon _{int}(T)\;,  \label{1}
\end{equation}
where $N$ is the number of observed pulse, $N_{0}$ is a constant, $\nu $ and 
$\dot{\nu}$ are the pulsar's rotational frequency and its time derivative, $%
\epsilon _{int}(T)$ is the pulsur's intrinsic noise, $T$ is the pulsar's
proper time, which is related to the observer's proper time $t$ by the
equation 
\begin{equation}
T=t+\Delta (T)+\Delta _{\odot }(t)+\epsilon _{pr}(t)\quad .  \label{2}
\end{equation}
Herein, $\Delta (T)$ and $\Delta _{\odot }(t)$ describe the classic and
relativistic time delay corrections for the binary system and for the Solar
system respectively, $\epsilon _{pr}(t)$ are noises produced by a number of
diverse perturbations during the time of propagation of pulses from the
pulsar to observer, inaccuracies in ephemerides of the solar system bodies,
imperfectness of atomic clocks used for the time metrology, and electronic
equipment used for timing observations. The equipment noise is performed to
be white irrespective of the longivity of the observations. On the other
hand, any other kind of noise present has low-frequency components which
show up when the measurements are carried out over a sufficiently long time
span. It is worth noting that standard procedures of data processing were
worked out only for white noise which has a gaussian distribution of timing
residuals. If timing residuals are dominated by red noise, the standard
statistical estimations give unrealistic (overestimated) numerical values of
the measured parameters.

\section{Red Noise and Limits on Estimates of Parameters}

Recently, we have begun to study the problem of pulsar data processing in
the presence of red noise (Kopeikin 1997, Kopeikin 1999a, Kopeikin \&
Potapov 1998). To model the red noise we have used a shot-noise
approximation with a specific choice of step function in such a way that any
rational spectrum of the red noise could be restored, including flicker
noise of phase ($~1/f$), random walk of phase ($~1/f^{2}$), and so on. The
noise model includes both stationary and non-stationary parts of the
autocovariance function which are well separated algebraically. In addition,
an exhaustive treatment of the polynomial drift of the noise was worked out
in full detail. Applying the model for processing fake data of a binary
pulsar in a circular orbit we set upper limits on the numerical values of
the parameter's variances. These upper limits depend on the total span of
observations $\tau $ and can either decrease or grow as $\tau $ increases.
The time dependence some of the limits are shown in Table 1.

\begin{table}[tbp]
\caption{ Dependence of variances of the pulsar's rotational frequency $\nu $
and some of the Keplerian and post-Keplerian parameters from the total span of
observations $\tau $ in case of presence of red noise with the spectrum $%
S(f)={\rm h}_{n}/f^{n}$, where ${\rm h}_{0}$, ${\rm h}_{1}$,...,${\rm h}_{6}$
are constants characterizing the intensity of the noise}
\begin{tabular}{l|llllll|}
&  &  &  &  &  &  \\ 
S(f) & $\quad \nu \quad \quad $ & $T_{0}\quad \quad $ & $x\quad \quad $ & $%
P_{b}\quad \quad $ & $\dot{x}\quad \quad $ & $\dot{P}_{b}\quad \quad $ \\ 
\hline
&  &  &  &  &  &  \\ 
${\rm h}_{0}$ & $\quad \tau ^{-3}$ & $\tau ^{-1}$ & $\tau ^{-1}$ & $\tau
^{-3}$ & $\tau ^{-3}$ & $\tau ^{-5}$ \\ 
&  &  &  &  &  &  \\ 
${\rm h}_{1}/f$ & $\quad \tau ^{-2}$ & $\tau ^{-1}$ & $\tau ^{-1}$ & $\tau
^{-3}$ & $\tau ^{-3}$ & $\tau ^{-5}$ \\ 
&  &  &  &  &  &  \\ 
${\rm h}_{2}/f^{2}$ & $\quad \tau ^{-1}$ & $\tau ^{-1}$ & $\tau ^{-1}$ & $%
\tau ^{-3}$ & $\tau ^{-3}$ & $\tau ^{-5}$ \\ 
&  &  &  &  &  &  \\ 
${\rm h}_{3}/f^{3}$ & const. & const. & const. & $\tau ^{-2}$ & $\tau ^{-2}$
& $\tau ^{-4}$ \\ 
&  &  &  &  &  &  \\ 
${\rm h}_{4}/f^{4}$ & $\quad \tau $ & $\tau $ & $\tau $ & $\tau ^{-1}$ & $%
\tau ^{-1}$ & $\tau ^{-3}$ \\ 
&  &  &  &  &  &  \\ 
${\rm h}_{5}/f^{5}$ & $\quad \tau ^{2}$ & $\tau ^{2}$ & $\tau ^{2}$ & const.
& const. & $\tau ^{-2}$ \\ 
&  &  &  &  &  &  \\ 
${\rm h}_{6}/f^{6}$ & $\quad \tau ^{3}$ & $\tau ^{3}$ & $\tau ^{3}$ & $\tau $
& $\tau $ & $\tau ^{-1}$ \\ 
&  &  &  &  &  & 
\end{tabular}
\end{table}
One can see that in the case when the spectral index $n=0,1,2,...$ of the
noise is big enough variances of some, or even all, parameters grow such
that one can not get improvements in testing GR, which prevents better
determination of masses of neutron stars and other physical characteristics
of the binary system. Special methods of observations and/or data processing
should be suggested to overcome this difficulty.

This problem can also be considered from a different point of view. The fact
is that some of the red noises have a specific astrophysical origin and
their study would deliver extremely valuable information about physical
processes generating such noises\footnote{%
A striking example which comes to mind is the case of electromagnetic cosmic
background radiation (CMB) which was discovered by Pensias \& Wilson as an
excess noise in the equipment they used for radio survey of the sky.}. One
particular example represents a low-frequency timing noise produced by the
variable gravitational fields of binary stars.

\section{Gravitational timing noise from binary stars}

Precise calculations of the autocovariance function of the gravitational
timing noise from binary stars requires having a mathematically complete
solution to the problem of propagation of electromagnetic waves in variable
gravitational fields of localized self-gravitating sources. Significant
progress in solving this problem has been achieved recently by Kopeikin et
al. (1999) and Kopeikin \& Sch\"{a}fer (1999). The relativistic time delay
in an arbitrary time-dependent gravitational field has been presented as a
function of the relative distances between observer, source of light, and
localized source of the non-stationary gravitational field as well as the
intrinsic characteristics of the source. Using the exprerssion for the time
delay, the gravitational timing noise from an ensemble of binaries in our
galaxy has been evaluated for the case of PSR B1937+21 under some
simplifying assumptions (Kopeikin 1999b).

Using the same approach the gravitational timing noise from an ensemble of
binaries in a globular cluster can also be parameterized and predicted.
Long-term precise timing monitoring of the bunch of millisecond pulsars in
47 TUC (or other globular cluster) will be required to test the presence of
the noise and its properties. It will help to better understand spatial
distribution, mass function, and other statistical properties of binaries in
the globular cluster.

\section{Acknowledgement}

We thank D. Moran for careful reading of the manuscript and useful comments.

\end{document}